\documentclass[sigconf]{acmart}
\usepackage{subcaption}
\usepackage[ruled]{algorithm2e}
\usepackage{multirow}
\usepackage{booktabs}

\copyrightyear{2023}
\acmYear{2023}
\setcopyright{acmlicensed}\acmConference[WWW '23]{Proceedings of the ACM
Web Conference 2023}{May 1--5, 2023}{Austin, TX, USA}
\acmBooktitle{Proceedings of the ACM Web Conference 2023 (WWW '23), May
1--5, 2023, Austin, TX, USA}
\acmPrice{15.00}
\acmDOI{10.1145/3543507.3583200}
\acmISBN{978-1-4503-9416-1/23/04}

\begin{document}

\title{Efficient and Low Overhead Website Fingerprinting Attacks and Defenses based on TCP/IP Traffic}
%
\author{Guodong Huang}
\authornote{Both authors contributed equally to this research. Corresponding authors: Chunpeng Ge and Zhe Liu.}
\affiliation{
  \institution{Nanjing University of Science and Technology}
  \city{Nanjing}
  \state{Jiangsu}
  \country{China}
}
\email{huangguodong@njust.edu.cn}

\author{Chuan Ma}
\authornotemark[1]
\affiliation{
  \institution{Zhejiang Lab}
  \city{Hangzhou}
  \state{Zhejiang}
  \country{China}
}
\email{chuan.ma@zhejianglab.edu.cn}

\author{Ming Ding}
\affiliation{%
  \institution{Data 61, CSIRO, Sydney}
  \city{Sydney}
  \country{Australia}}
\email{ming.ding@data61.csiro.au}

\author{Yuwen Qian}
\affiliation{%
  \institution{Nanjing University of Science and Technology}
  \city{Nanjing}
  \state{Jiangsu}
  \country{China}
}
\email{admon@njust.edu.cn}

\author{Chunpeng Ge}
\affiliation{
 \institution{Shandong University.}
 \city{Jinan}
 \state{Shandong}
 \country{China}
}
\email{gechunpeng@126.com}

\author{Liming Fang}
\affiliation{%
  \institution{Nanjing University of Aeronautics and Astronautics Shenzhen Research Institute, Nanjing University of Aeronautics and Astronautics.}
  \city{Nanjing}
  \state{Jiangsu}
  \country{China}
}
\email{fangliming@nuaa.edu.cn}

\author{Zhe Liu}
\affiliation{%
  \institution{Zhejiang Lab}
  \city{Hangzhou}
  \state{Zhejiang}
  \country{China}
}
\email{zhe.liu@zhejianglab.com}

\begin{abstract}
  Website fingerprinting attack is an extensively studied technique used in a web browser to analyze traffic patterns and thus infer confidential information about users. Several website fingerprinting attacks based on machine learning and deep learning tend to use the most typical features to achieve a satisfactory performance of attacking rate. However, these attacks suffer from several practical implementation factors, such as a skillfully pre-processing step or a clean dataset. To defend against such attacks, random packet defense (RPD) with a high cost of excessive network overhead is usually applied. In this work, we first propose a practical filter-assisted attack against RPD, which can filter out the injected noises using the statistical characteristics of TCP/IP traffic. Then, we propose a list-assisted defensive mechanism to defend the proposed attack method. To achieve a configurable trade-off between the defense and the network overhead, we further improve the list-based defense by a traffic splitting mechanism, which can combat the mentioned attacks as well as save a considerable amount of network overhead. In the experiments, we collect real-life traffic patterns using three mainstream browsers, i.e., Microsoft Edge, Google Chrome, and Mozilla Firefox, and extensive results conducted on the closed and open-world datasets show the effectiveness of the proposed algorithms in terms of defense accuracy and network efficiency.
\end{abstract}



\maketitle

\section{Introduction}
With the overgrowing Internet and rapid developments of web browser techniques, the web browser has become an essential tool in people's daily life. When a user visits websites, such as online shopping, social networking, online banking, etc., the destination will be inevitably leaked along a number of routers. These routers may accidentally observe or intentionally collect users' behaviors, and further infer sensitive information that violates individuals' privacy with an increasing severity~\cite{wang2013improved}. To prevent such privacy leakage, the Tor browser was developed based on The Onion Routing technology to enable anonymous communication, which is designed to conceal the identity of the users by encryption in the application layer of the communication protocol stacks~\cite{dingledine2004tor}. However, an attacker can still attempt to compromise the user's information by observing the patterns in the sequence of packets, e.g., by means of website fingerprinting (WF) attacks~\cite{cai2014systematic,chen2010side,murdoch2005low,sirinam2018deep}.

WF attack is proposed to reveal a user's browsing behavior by analyzing the exposed traffic patterns, even if they are encrypted~\cite{cai2012touching}. For example, handcrafted features, such as order, direction, size, and length of each connection, are extracted to represent Tor traffic and fed to classifying algorithms, such as support vector machine (SVM)~\cite{panchenko2016website}, random forest~\cite{wang2014effective}, and k-nearest neighbors (k-NN)~\cite{hayes2016k}, to launch attacks. However, such attacking methods require massive pre-processing operations based on expert knowledge and will lose efficiency under dummy-based defensive algorithms. For example, the adaptive padding~\cite{perry2015padding} and its follow-up, WTF-PAD~\cite{juarez2016toward}, have been shown the effectiveness against the attacks based on handcrafted features with reasonable overheads, and have been adopted in Tor. Furthermore, Wang and Goldberg proposed Walkie-Talkie~\cite{203876}, a low overhead defensive algorithm, to beat such attack algorithms, and ease the implementation by automated extraction of traffic features. On the contrary to defense, deep learning-based attacking algorithms have been proposed and become increasingly popular~\cite{sirinam2018deep,Rimmer_2018,sirinam2019triplet}.

In~\cite{sirinam2018deep}, a deep fingerprinting attack using convolutional neural networks was proposed, and significant performance gains were achieved in both closed and open-world scenarios. In detail, a nearly 99$\%$ accuracy has been achieved by only using the direction of the packets, following Wang \emph{et al.}'s methodology~\cite{wang2014effective}. To survive such deep learning attacks, an effective but expensive mechanism is to add random perturbations to the traffic packets. Recently, an explainable artificial intelligence (XAI) based algorithm was proposed in~\cite{9556572} and analyzed the leakage points to design a high-efficiency traffic perturbation method.

\begin{figure}[t]
  \centerline{\includegraphics[width=0.35\textwidth, height=3cm]{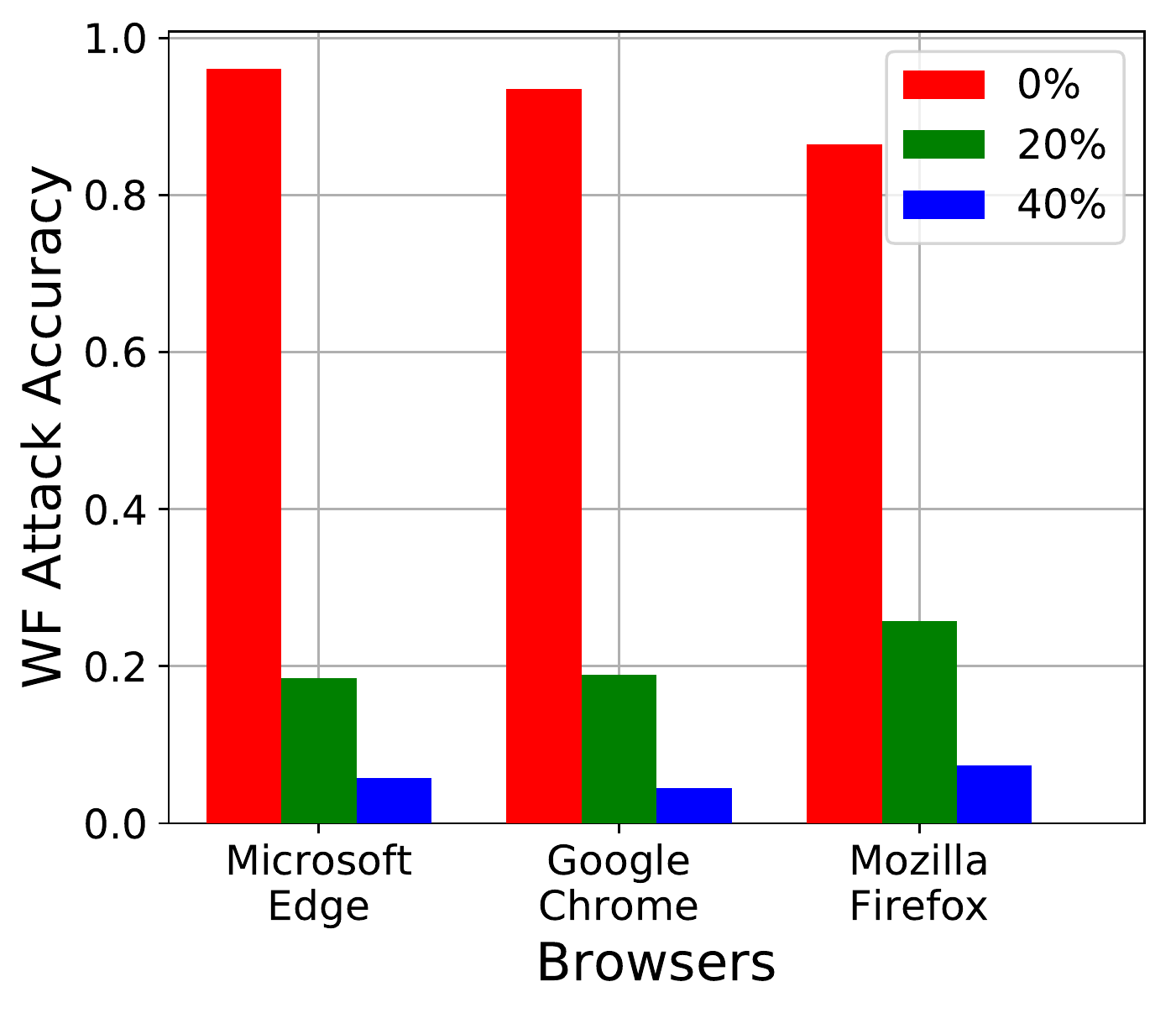}}
  \caption{WF Attack Accuracy under Random Noise Packets}
  \label{wf_noise}
\end{figure}

All of these methods are focusing on the important link of network traffic. When collecting and processing the Tor network traffic, the adversary can choose to directly capture TCP/IP packet instances, so that there will be two features of packet size and packet direction (e.g., 544, -1088, 1088, where these numbers represent bytes
and that the sign represents the direction of the data flow) in each traffic instance, or parse TCP/IP traffic to obtain Tor cell with more consistent packet size (i.e., 512 bytes for each cell), which results in only one feature of packet direction (e.g., 1, -1, -1, 1, 1). However, even though Tor cells can conceal the feature of packet size after parsing, one or more Tor cells will still be converted into TCP/IP instances with packet size in the actual network transmission~\cite{wang2013improved}. Furthermore, in the face of the most common defense of adding random noise packets, WF attacks will not have a good solution. For example, as shown in Figure~\ref{wf_noise}, we evaluate the accuracy of the WF attack when adding random noise packets, and the results show that when the noise probability is greater than 20\%, the attack performance will be greatly reduced. Therefore, an intriguing question arises, what if the attacker can utilize the information on packet size and design an advanced attacking method to identify the added packets?

Motivated by this, in this paper, we reconsider the attacking and defense methods in WF. In detail, we take the traffic packet size as well as direction into account, and first design a filter-based attacking method to differentiate the real and perturbed traffic packets and show the effectiveness of the method using real-life datasets collected via three mainstream web browsers. To defend against such an attack, we further propose a list-based defensive mechanism, which can mitigate the proposed attack and achieve high efficiency in terms of transmission overhead. Specifically, our main contributions are listed as follows.
\begin{itemize}
  \item We first illustrate a detailed example of how the random traffic packets-based defensive mechanisms can be used to combat the WF attack when considering the factors of packet size.
  \item To fully utilize the benefits of packet size, we then from a viewpoint of the attacker, propose a filter-assisted attack.
  \item To complete our study, a list-based defensive mechanism is designed to defend against such attacks. In addition, to further relax the transmission overhead, a list-based splitting defensive algorithm is proposed, which can adjust the configurable tradeoff between the defense rate and the injected overhead.
  \item We show details on collecting real traffic packets via three mainstream web browsers to construct the experimental datasets.\footnote{Related codes can be found in https://github.com/guduin/wf} Extensive experiments are conducted to show the effectiveness of the proposed attacking as well as the defensive mechanisms.
\end{itemize}

\section{Background}

Website fingerprinting attack, as a new attack technique for de-anonymization, has been proven to be effective in identifying websites. To defend against such attacks, amounts of defense algorithms have been proposed, which mainly attempt to cover the unique or routine traffic distributions by adding dummy packets. In the process of attack and defense, the direction, size, length, and time of traffic can be used as extracted features. Many studies have tended to use the characteristics of packet direction and size while others only use the packet direction.

\subsection{Methods with Packet Direction and Size}

Cui \emph{et al.} in ~\cite{9355590} evaluated the WF attack method and extended it to multiple web page scenarios. They used the TCP layer packets directly (including packet sizes and direction of network packets). Applying the same features, Yin \emph{et al.}~\cite{9514394} studied the problem of overlapping noise in network traffic extraction, and proposed an automated multi-tab WF attack method. When a user opens multiple pages consecutively in a short time, the overlapping of subsequent traffic will degrade the performance of traditional WF attacks, and the proposed method can filter out the subsequent overlapping traffic.

In addition, Zuo \emph{et al.}~\cite{8067534} proposed a novel method for WF attack, which combines website fingerprinting and DNA-based fingerprints in biology to design attack models with the help of the profile hidden Markov model (PHMM). They used the size and direction of packets and symbolized them with two indicators as features. Shusterman \emph{et al.} in~\cite{9072556} proposed a novel traffic acquisition method by monitoring the cache occupancy channel instead of the traditional traffic patterns. They monitored the cache size by sending JavaScript codes to the target's router, achieving more resistance to network-based fingerprinting attacks.

In the study of defense methods, the first work based on this strategy, BuFLO, was proposed in~\cite{6234422}, but it failed to conceal coarse features, such as time duration, size, and the total length, unless with intolerable overheads. Furthermore, Gulmezoglu \emph{et al.}~\cite{9556572} proposed a defensive approach based on XAI, which can analyze and obtain leak points in ML classification tasks and then perturb these points to obscure the traffic features. Similar to Shusterman~\cite{9072556}, the authors in~\cite{9556572} used the data collected by the cache side channel. Eric \emph{et al.}~\cite{8416437} proposed a defense method based on clustering, which merges a group of similar websites into an indistinguishable maximum set, and finally achieved a relatively high defense rate at an expensive cost on transmission overhead.

Through these studies, we can find that although network traffic has many characteristics, most studies choose packet size and packet direction, or the cache side channel that can be characterized as the data source.

\subsection{Methods with Packet Direction Only}
Extracting more traffic features means that more information should be inferred in advance. However, processing more information also occupies more resources. If multiple features are not properly integrated or non-typical features are incorporated, the performance of WF will be degraded. Therefore, there are also many scholars who have made another drop in packet size and packet direction. Only using the least features (i.e., package direction) has become a popular topic~\cite{wang2014effective, sirinam2018deep, Rimmer_2018, he2018deep}.

Since Wang \emph{et al.}~\cite{wang2014effective} have proved that only using packet directions can also achieve the topmost accuracy, more and more research tends to use package direction solely. For example, Sirinam \emph{et al.}~\cite{sirinam2018deep} proposed a deep fingerprinting (DF) attack technology based on a convolutional neural network (CNN). Dazzing results, more than 98$\%$ accuracy, have been achieved based on Tor traffic. In addition to WTF-PAD~\cite{juarez2016toward} and intercom defense measures~\cite{203876}, obvious performance gains are also obtained. Rimmer \emph{et al.} in~\cite{Rimmer_2018} followed the direction-only network traffic and collected a dataset comprised of more than three million network traces and found that the performance achieved was comparable to previous studies. Meanwhile, the implicit features automatically learned through their approaches are more resilient to the dynamic changes of web content over time. After that, He \emph{et al.} ~\cite{he2018deep} used Rimmer's datasets and DL, achieving more than 99\% accuracy. Their experimental results show that deep learning is an efficient and robust fingerprinting attack technique for websites.

\begin{figure}
  \centering
  \includegraphics[width=0.45\textwidth,height=4cm]{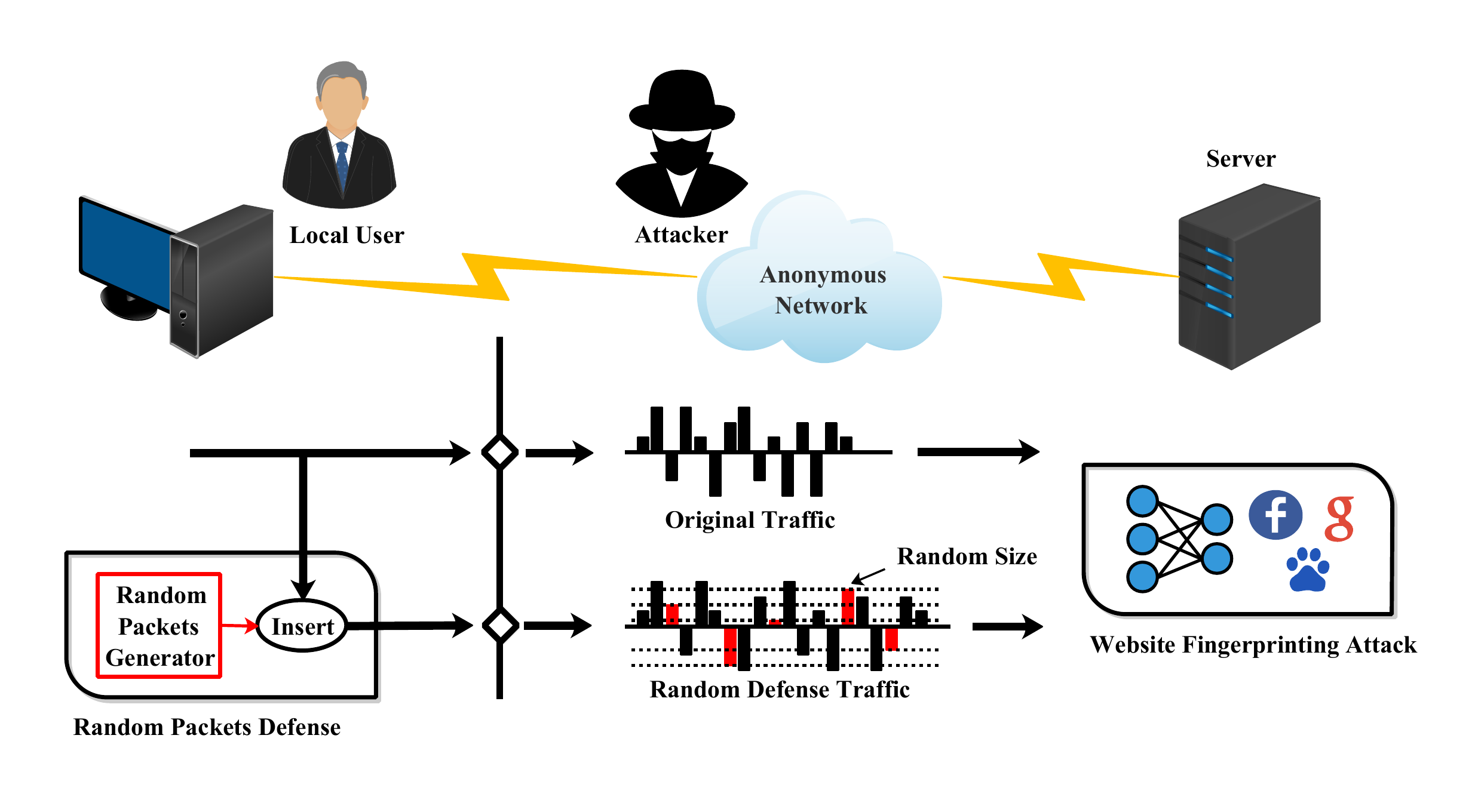}
  \caption{System Model}\label{system_model}
\end{figure}

The above researches show that only using packet direction can achieve good attacking results. However, the limited information will degrade the attacking capability on the awareness of the traffic. In addition, the aforementioned algorithms especially rely on clean datasets, which will suffer from adding random noises directly (such as the RPD algorithm in Figure.~\ref{system_model}). Therefore, in this work, we are more inclined to use more information to assist the attack in order to achieve stronger robustness and anti-interference.

\section{Overview}
\subsection{Adversary Model}
Although the aim of Tor is to protect users against adversaries from distinguishing which site the client is transiting. WF attacks, basically based on deep learning techniques, can undermine this protection by analyzing traffic patterns. To deploy such an attack, adversaries first need to capture the traffic patterns from the current website to the destination, and extract unique features for each transition of each website. Several features, such as transmission time and size with directions~\cite{panchenko2011website,wang2014effective,sirinam2018deep}, packet length and its frequencies~\cite{herrmann2009website}, distance-based scores~\cite{wang2013improved}, and the number of bursts~\cite{wang2014effective}, is then used to train a supervised classifier. Finally, the adversary can collect new transitions via Tor or other mainstream web browsers, and launch WF attacks.

In this work, we follow the assumption in~\cite{sirinam2018deep} that adversaries can only have access to the link between the user and the connected network within the used web browser and capture the traffic packets between links. As mentioned in~\cite{sirinam2018deep}, positions that a potential adversary launch WF attacks include: eavesdroppers on the client's local network, local system administrators, internet service providers (ISP), autonomous systems (AS) between the client and the entry network, and the operators of the entry network. In Figure~\ref{system_model}, we illustrate the adversary scenario: the user surfs the Internet by clicking links via Tor or other anonymous networks, and the adversary intents to extract the destination of each click by analyzing the traffic pattern between the client and the anonymous network. To this end, an adversary first accesses a list of potential websites through Tor or other web browsers and collects traffic traces between the local user and destinations. Then, traffic traces are labeled as their corresponding websites, and classified by the unique features (e.g., packet length, traffic direction, and traffic size) using supervised learning methods (e.g., random forest, k-NN, and SVM) or DL-based methods (e.g., SDAE, LSTM, and CNN).

\subsection{A Toy Example of Random Packets Defense}
In this subsection, we, from a defender viewpoint, first illustrate a basic process of RPD. As shown in Figure~\ref{system_model}, two features (packet size and direction) are prominent during the collection. To degrade the attack performance, a direct way is to disturb the distributions of these features by adding random packets. The execution logic of the random packets generator is as follows:

\begin{equation}
P_r = (-1)^{RANDI(1, 2)} \times RANDI(S_m, S_M) {\rm ,}
\end{equation}

\noindent that is, output a packet $P_r$ with a random direction and size, where the size is a random integer between the minimum size $S_m$ and the maximum size $S_M$. Then, the defender can process Algorithm~\ref{random} to start the defense. In detail, by inserting random traffic packets into the original trace, the internal connections between features are interrupted. During each perturbation, two factors, the injected position and the injected traffic size with direction are considered. For example, for each traffic bar, the defender set a random probability $P_t$ to determine whether there should be inserted a perturbation, and a random size of the traffic bar will be inserted when a random number ([0, 1]) is smaller than $P_t$.

In practice, RPD is easy to implement and deploy. However, from the statistical characteristics of packets, the inserted random packets will be significantly different from the original traffic. If the attacker takes advantage of this vulnerability, RPD may be disabled. Therefore, a high effective and comprehensive defensive mechanism should be proposed.
\begin{algorithm}[t]
  \small
	\caption{Random Packets Defense (RPD) Algorithm}
  \label{random}
	\LinesNumbered
	\KwIn{Original traffic vector $V_{ORG}$}
	\KwOut{Random Packets Defense traffic vector $V_{RPD}$}
	\ForEach{$Packet$ in $V_{ORG}$}{
        Output $Packet$ to $V_{RPD}$\\
		\If{random number \textless $P_t$}{
            Generate a $Random Packet$\\
            Output $Random Packet$ to $V_{RPD}$\\
		}
	}
\end{algorithm}

\begin{figure*}[t]
  \centering
  \begin{subfigure}{0.25\linewidth}
  	\centering
  	\includegraphics[width=0.9\linewidth, height=3cm]{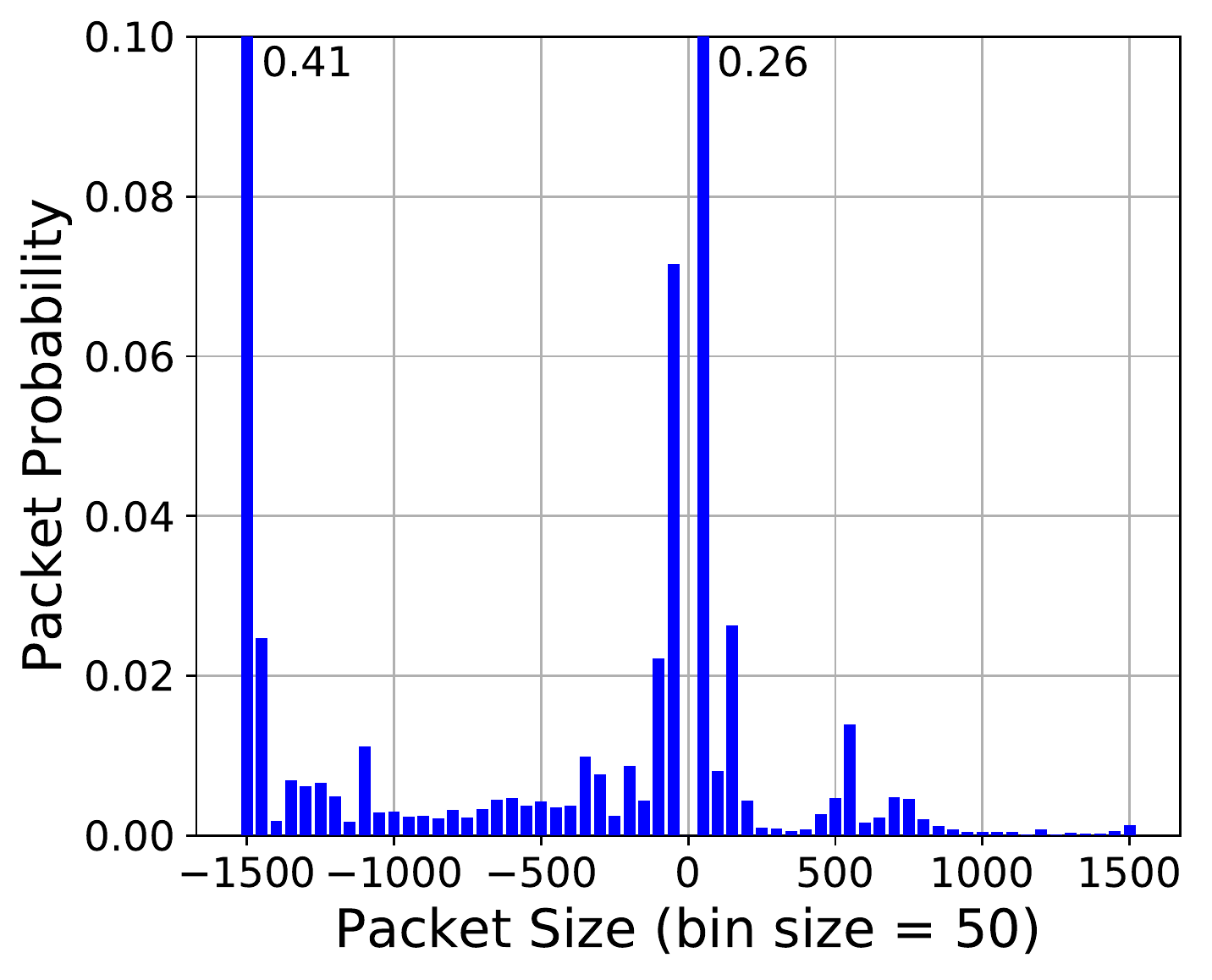}
  	\caption{Microsoft Edge}
  \end{subfigure}
  \begin{subfigure}{0.25\linewidth}
  	\centering
  	\includegraphics[width=0.9\linewidth,, height=3cm]{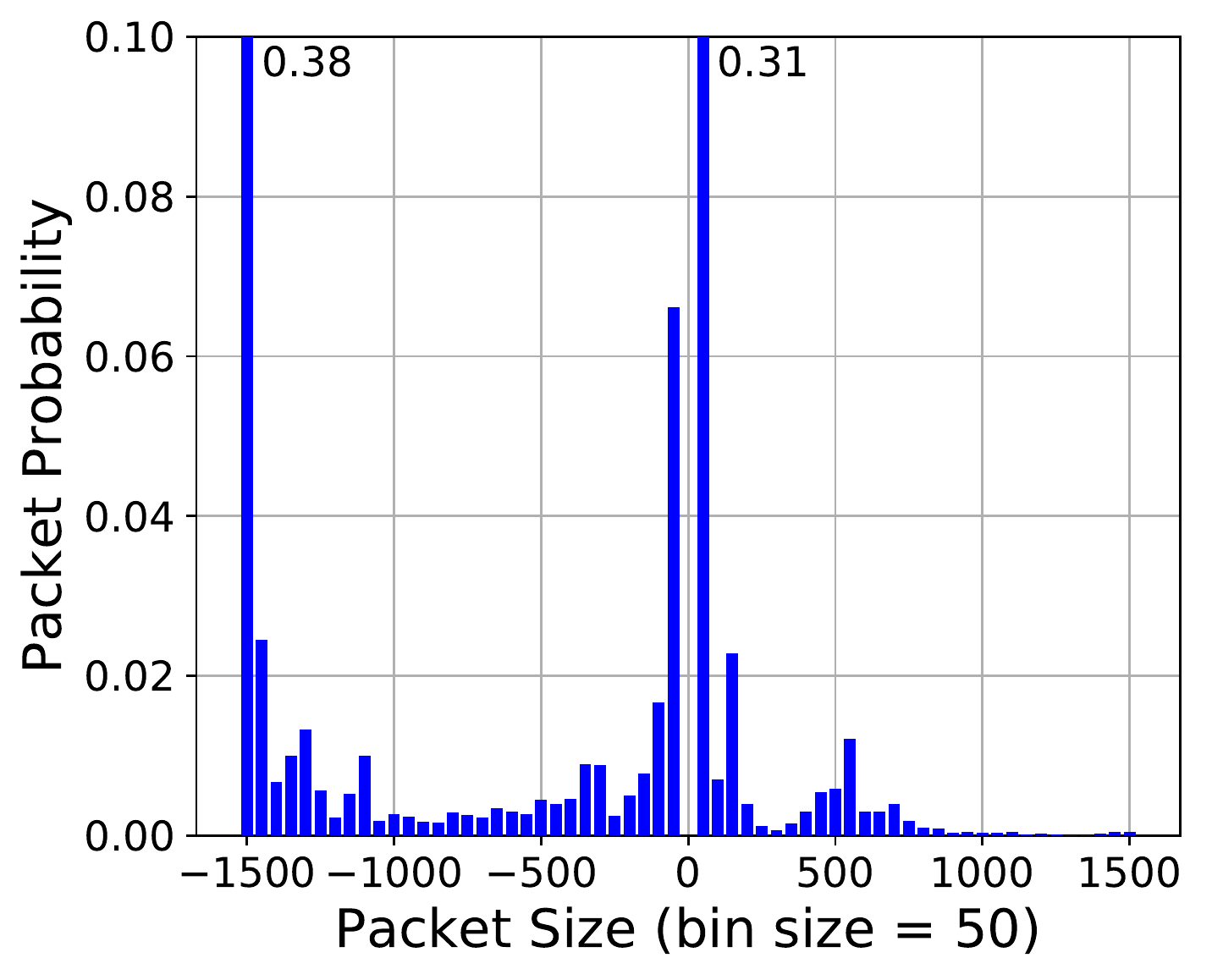}
  	\caption{Google Chrome}
  \end{subfigure}
  \begin{subfigure}{0.25\linewidth}
  	\centering
  	\includegraphics[width=0.9\linewidth,, height=3cm]{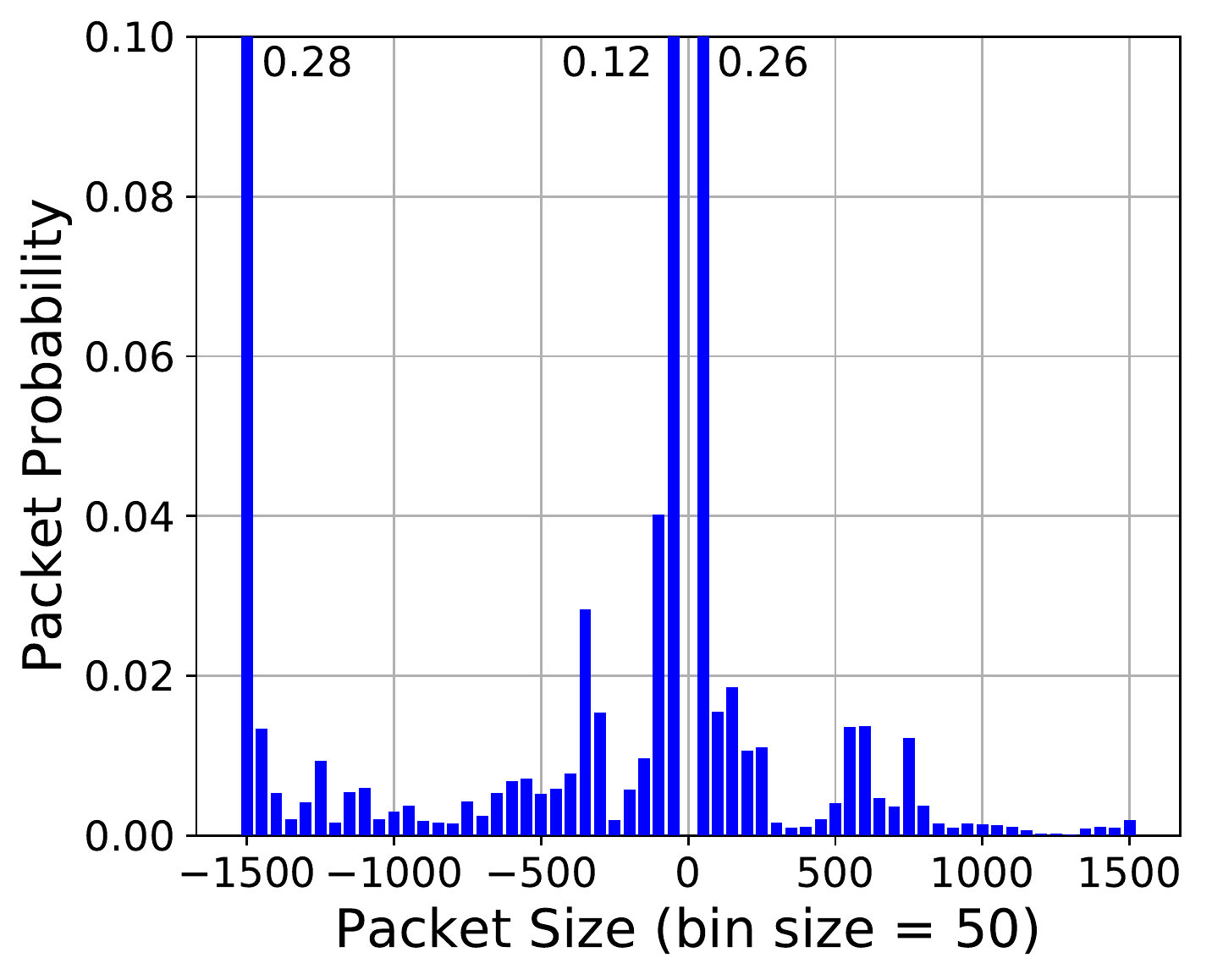}
  	\caption{Mozilla Firefox}
  \end{subfigure}
  \caption{Packet Size Distribution of Different Browsers}
  \label{size}
\end{figure*}

\section{Proposed Attack and Defense Method}

In this section, we will first analyze the shortcomings of RPD, then propose an enhanced attack method to break RPD. Finally, a novel defense method has been proposed to deal with the enhanced attack.

\subsection{Filter Assisted Attack}

Destroying traffic characteristics is the main principle of anti-WF attacks. As discussed in the previous chapter, adding noise packets randomly may be a straightforward and efficient method. In order to observe the packet size distribution of different browsers, we have captured 8000 traffic from 100 websites in three main web browsers, e.g., Microsoft Edge, Google Chrome, and Mozilla Firefox. Afterward, we calculate the relationship between each packet size and the occurrence probability, illustrated in Figure~\ref{size}. The abscissa is the packet size (outbound packets are defined as positive, and inbound packets are defined as negative), and the ordinate is the probability of the corresponding packet size. We can clearly find that the distribution of package sizes is extremely unbalanced. Therefore, if these statistical characteristics are not taken into account during the attack, newly inserted interference packets will be easily filtered out.

\begin{algorithm}[t]
  \small
	\caption{Filter Assisted Attack (FAA) Algorithm}
  \label{filter}
	\LinesNumbered
	\KwIn{Lots of original traffic vector $V_{LORG}$}
	\KwIn{Random Packet Defense traffic vector $V_{RPD}$}
	\KwOut{Filter Assisted Attack traffic vector $V_{FAA}$}

    $Statistics$ of $V_{LORG}$ with $S_p$\\
    $Sort$ of $Statistics$ in descending by $P_p$\\
    $L$ gets the first $X$ packets of $Sort$

	\ForEach{$Packet$ in $V_{RPD}$}{
		\If{$Packet$ Size not in $L$}{
			continue
		}
        Output $Packet$ to $V_{FAA}$\\
	}
\end{algorithm}

\begin{figure}[t]
  \centerline{\includegraphics[width=0.4\textwidth]{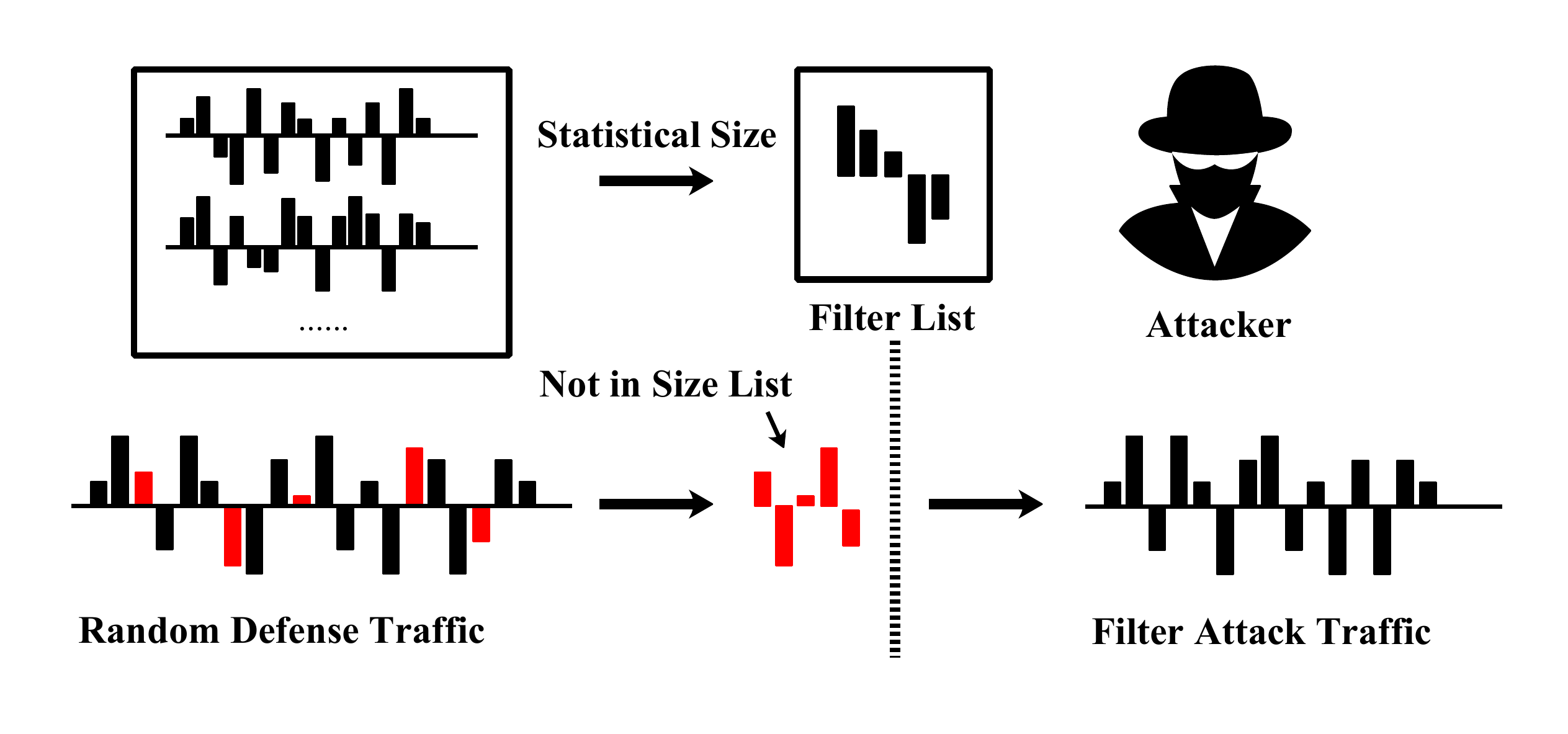}}
  \caption{Random Packets Defense (RPD) through Filter Assisted Attack (FAA)}
  \label{traffic_filter}
\end{figure}

Based on the above observation, we have designed an enhanced attack method, called filter-assisted attack (FAA), to destroy the RPD. Its algorithm is shown in Algorithm~\ref{filter}. The main idea is as follows:
\begin{itemize}
  \item \textbf{Step 1.} (Line 1-3) Analyze lots of network traffic $V_{LORG}$ and make a statistical analysis on the size of all packets. Obtain the size of the packet with the highest frequency, and then take the first $X$ as the filter list $L$.
  \item \textbf{Step 2.} (Line 4-9) Filtrate the traffic of random packets defense $V_{RPD}$, and the packets whose sizes are not in the filter list will be filtered out. Then, output the filtered traffic $V_{FAA}$.
\end{itemize}
Its working model is shown in Figure~\ref{traffic_filter}. Due to the packet size difference, randomly added noises are filtered out by the designed filter.
Because the packet of random noise is evenly distributed, there will be a lot of interferences distributed out of the filter list. So these disturbances will be filtered out and the FAA can improve the adaptability to random noises. Although the filter can not completely filtrate the injected random noise, the overall noise level has been reduced.

\subsection{List Assisted Defense}

\begin{algorithm}[t]
  \small
	\caption{List Assisted Defense (LAD) Algorithm}
  \label{list}
	\LinesNumbered
	\KwIn{Lots of original traffic vector $V_{LORG}$}
	\KwIn{Original traffic vector $V_{ORG}$}
	\KwOut{List Assisted Defense traffic vector $V_{LAD}$}

    $Statistics$ of $V_{LORG}$ whit $S_p$\\
    $Sort$ of $Statistics$ in descending by $P_p$\\
    $L$ gets the first $X$ packets of $Sort$\\
    $P_L$ gets $P_p$ of $L$

    \If{base method = insert}{
    	\ForEach{$Packet$ in $V_{ORG}$}{
            Output $Packet$ to $V_{LAD}$\\
    		\If{random number \textless $P_t$}{
                Select $List Packet$ from $L$ according to $P_L$\\
                Output $List Packet$ to $V_{LAD}$\\
    		}
    	}
    }
    \If{base method = split}{
        \ForEach{$Packet$ in $V_{ORG}$}{
    		\If{random number \textless $P_t$}{
                Select $List Packet$ from $L$ according to $P_L$\\
                \If{$Packet$ - $List Packet$ $\geq$ $S_m$}{
                    Output $List Packet$ to $V_{LAD}$\\
                    Output $Packet$ - $List Packet$ to $V_{LAD}$\\
                }
    		}
    	}
    }

\end{algorithm}

\begin{figure}[t]
  \centerline{\includegraphics[width=0.35\textwidth]{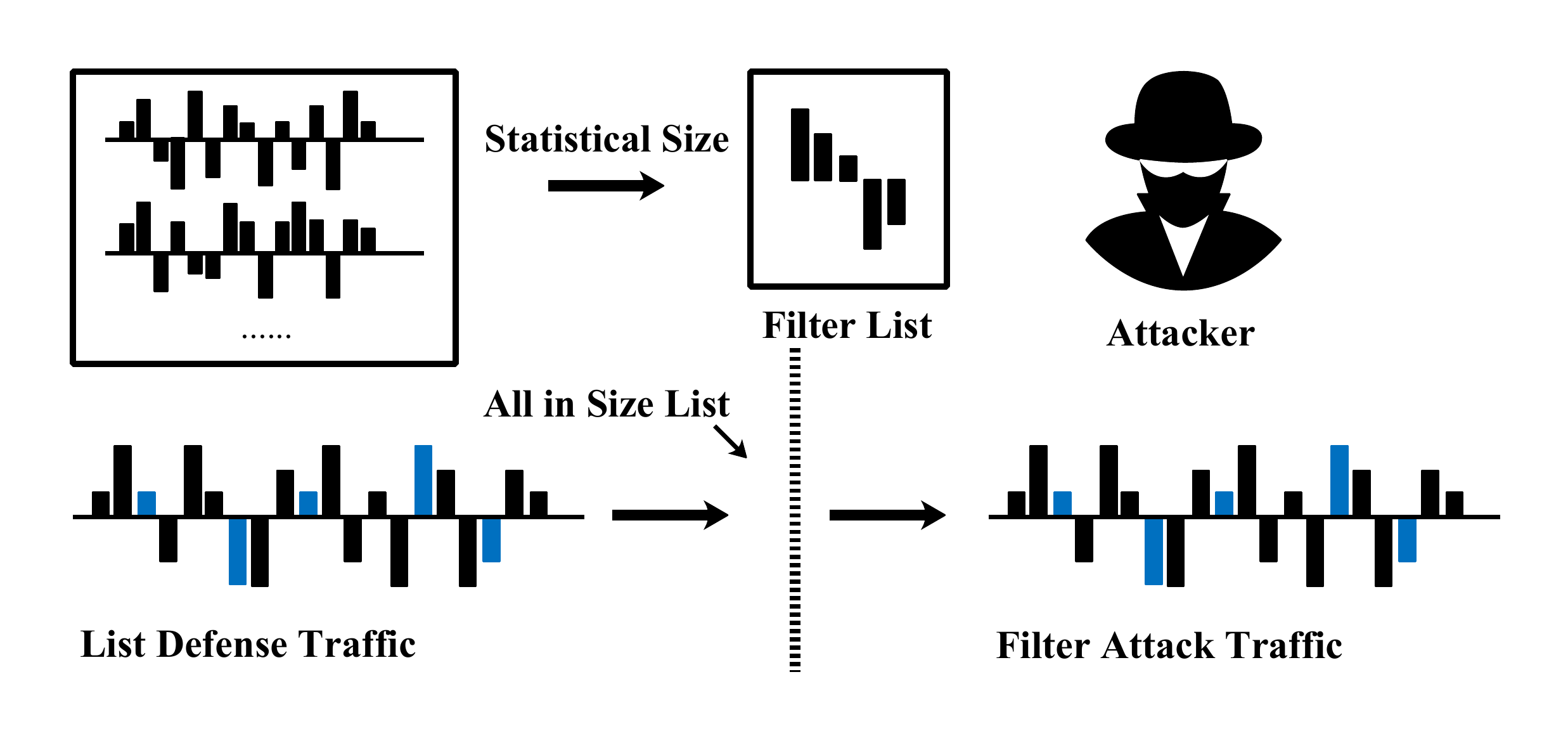}}
  \caption{List Assisted Defense (LAD) through Filter Assisted Attack (FAA)}
  \label{traffic_list}
\end{figure}

The packet size of RPD deviates from the distribution of normal packet size, so random injected patterns are easy to be filtered directly. Therefore, against such attacks, we further propose the list-assisted defense (LAD). Meanwhile, as the network overhead should also be carefully designed, we then provide two methods, one is based on probably packet insertion, and another is based on probably packet splitting, which has the purpose of alleviating the overloaded transmission overhead.
Its algorithm is as shown in Algorithm~\ref{list}. The main idea of this defense method is as follows:
\begin{itemize}
  \item \textbf{Step 1.} (Line 1-4) This part is similar to FAA. But the defender will obtain a list of probability distributions $P_L$ from $L$ in addition. $P_L$ will guide the defender to add noise packets according to the original probability distribution of the traffic.
  \item \textbf{Step 2.} (Line 5-13) If the defender chooses the packets insertion method, a listing packet $P_l$ will be appended with a determined probability after the output of an original packet. This list packet is guided by $P_L$ and taken from $L$ to ensure that its size and probability are consistent with the original traffic. The operation proceeds as the following equation:
      \begin{equation}
      P_l = {\rm Array}[L \times P_L]_{RANDI(0, N)} {\rm .}
      \end{equation}
  \item \textbf{Step 3.} (Line 14-24) If the defender chooses the packets splitting method, a large-size of traffic packet will be split into two smaller packets, and these newly generated packets are selected from the list. The newly generated first packet ($P_f$) and second packet ($P_s$) are processed as the following equation:
      \begin{equation}
      \begin{cases}
      P_f = P_l, P_s = P - P_l,& {\rm if\ } P_2 \geq S_m {\rm ;}\\
      P_f = P, P_s = {\rm None},& {\rm if\ } P_2 \textless S_m {\rm .}
      \end{cases}
      \end{equation}
\end{itemize}

When interfering, the defender tries to insert packets from $L$ with a probability $P_L$. In this way, it not only ensures the characteristics of random interference but also avoids filtering out due to the difference in statistical characteristics. Thus, when it passes through FAA again, limited noisy packets are filtered out. Its working model is shown in Figure~\ref{traffic_list}.

Inserting packets is easy to implement and does not affect normal communication, but because the inserted packets are usually useless noise, the overhead is huge. To reduce such overhead, packet splitting has the advantage that all packets are involved in the communication. According to our research in~\cite{socolofsky1991tcp}, each packet consists of a header and a payload, and the network itself splits the payload. If the defender controls this behavior, a variety of padding methods can be obtained, which will greatly increase the uncertainty and complexity of network traffic. So it is feasible to use this method to deal with WF attacks at a very limited extra cost on data packet headers.

\section{Data Collection}

\begin{figure}[t]
  \centerline{\includegraphics[width=0.45\textwidth,height=4cm]{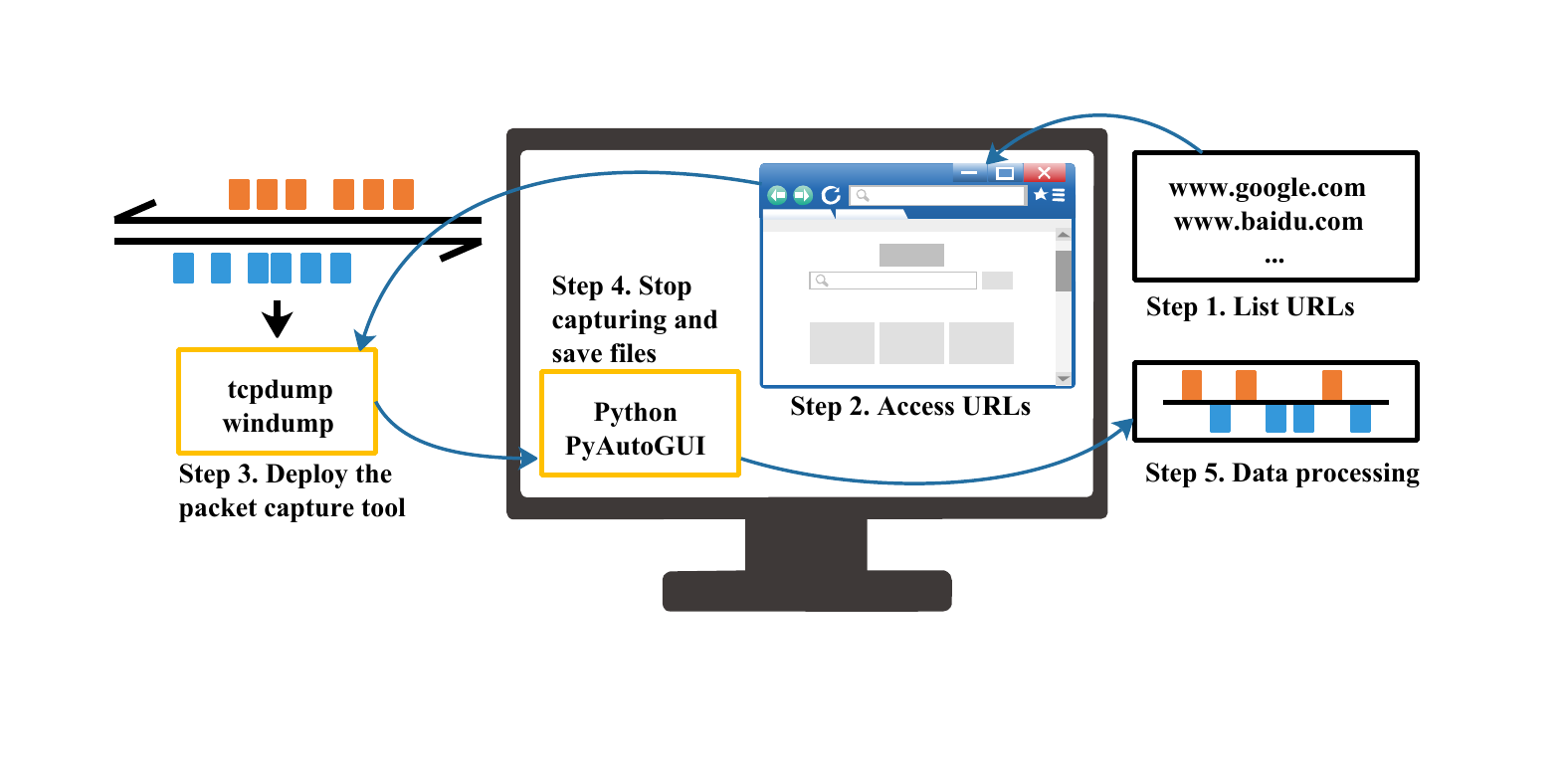}}
  \caption{Data Collection Flow Chart}
  \label{data_collection}
\end{figure}

In order to conduct model training and performance evaluation on WF attacks and defenses, it is essential to capture the traffic data of each website.
In this experiment, we use \emph{windump} for packet capture and Python and PyAutoGUI for automation. Its flow chart is shown in Figure~\ref{data_collection}, and the main steps are described as follows.

\begin{itemize}
  \item \textbf{Step 1:} list URLs. In order to traverse the website and capture traffic data, we first enumerate a list of website URLs. The common way to obtain the URL list is the traffic ranking of major websites. In the website traffic rankings \footnote{\href{https://top.chinaz.com/all/}{https://top.chinaz.com/all/}}, we can collect the URLs of the most popular sites as the target websites for subsequent capture.

  \item \textbf{Step 2:} access URLs. Once the URLs of the target websites are obtained, we can use Python to read the URL in turn and use PyAutoGUI to automatically type the URL into the browser for access. In addition, to ensure traffic consistency, we will disable the browser's cache for each instance.

  \item \textbf{Step 3:} deploy the packet capture tool. When a website is opening, we use \emph{windump} to capture the traffic packets. In order to filter out the interference of packets generated by other computer programs, we also need to set up a filter, achieved by using URL as a parameter.

  \item \textbf{Step 4:} stop capturing and saving files. After the site is loaded, we will stop this capture and save the file. To determine whether the website has finished loading, PyAutoGUI is used to monitor the pixels of the loading status icon on the screen. We also set a timeout for sites that have not finished loading. After capture, the observed packets will be stored as \emph{.pcap} files.

  \item \textbf{Step 5:} data processing. In order to carry out DL training, we need to process the \emph{.pcap} files. First, we use the \emph{dpkt} module in Python to parse the \emph{.pcap} file iteratively, and extract the size and direction of the packet. For the emitted packet, we define it as +size, and for the received packet, we define its data as -size. We end up with a bunch of vectors that represent per WF access. Considering that the length of each vector is different, we need to trim its length and finally unify it to a fixed length for the input of the DL model. Finally, we store the processed data as \emph{.csv} format.
\end{itemize}

So far, the data collection and processing work has been completed. In this experiment, we collected 100 URLs and repeat the visit 100 times, resulting in a 100*100 \emph{.pcap} files. After data pre-processing, we obtain the dataset with 100*100 data vectors.
For the open world, we have collected an additional 1000 URLs and visited every URL once, resulting in a 1000*1 dataset. In addition, in order to test the adaptability of WF in different browsers, we apply three main web browsers, Microsoft Edge, Google Chrome, and Mozilla Firefox, to process the above collection.

\section{Evaluation}

\subsection{Experimental Setup}

Our experiment is based on Windows 10 platform, the Microsoft Edge browser version is 105.0.1343.50, the Google Chrome browser version is 104.0.5112.102 and the Mozilla Firefox browser version is 105.0.1. We run our scripts and code in Anaconda 4.5.11, Jupyter Notebook with Python 3.7.0, and use a DL framework with TensorFlow as the back end and Keras as the front end.

In this experiment, we follow the work of Sirinam~\cite{sirinam2018deep}, Wang~\cite{wang2022snwf}, Rimmer~\cite{Rimmer_2018} \emph{et al.}, and take the CNN model as the experimental basis. Different from most two-dimensional image recognitions~\cite{jordan2015machine}, the input data of this experiment is only a one-dimensional data vector. Therefore, Conv1D and MaxPooling1D are used instead of Conv2D and MaxPooling2D and the size of the convolution kernel is changed from the classical (2, 2) to (5, 1). In the four times convolution - convolution - pooling unit, we gradually increase the number of convolution kernels to adopt more features. In the experiment, we use the number of convolution kernels of 32, 64, 128, and 256, respectively. After each layer of the network, we will add the activation function of ReLU to increase the nonlinearity of the model, which prevents the model from degrading to a linear programming structure. In the last layer of the model, the SoftMax function is used to connect the network to 100 classification result outputs. After the model is established, we use the categorical cross entropy compilation model and Adam's optimization model~\cite{sirinam2018deep}\cite{wang2022snwf}. For the collected data, we have divided them into the training dataset and test dataset according to the ratio of 8: 2, that is, 80 visits of 100 websites are used for the training model, and the remaining 20 visits are used for the testing model performance.

\subsection{Metrics}

\subsubsection{Defensive Efficiency}
 In order to comprehensively evaluate the defense effect,  we define the defense efficiency (DE) to represent the average of the classification accuracy with the probability of adding noise ranging from 0\% to 100\% that
${\rm DE} = \frac{N}{\sum_{i=0}^N{\rm Accurary}}{\rm .}$
\noindent where \emph{N} is the number of test points between 0\% to 100\%.

\subsubsection{Overhead}

The network cost of defense is a factor that cannot be ignored by any WF defensive method. Therefore, we define ${\rm Overhead} = \frac{\sum\ {\rm Extra\ Packets\ Size}}{\sum\ {\rm Original\ Packets\ Size}}{\rm .}$ to evaluate the overhead of defense.

\begin{figure*}[t]
	\centering
	\begin{subfigure}{0.24\linewidth}
		\centering
		\includegraphics[width=0.9\linewidth]{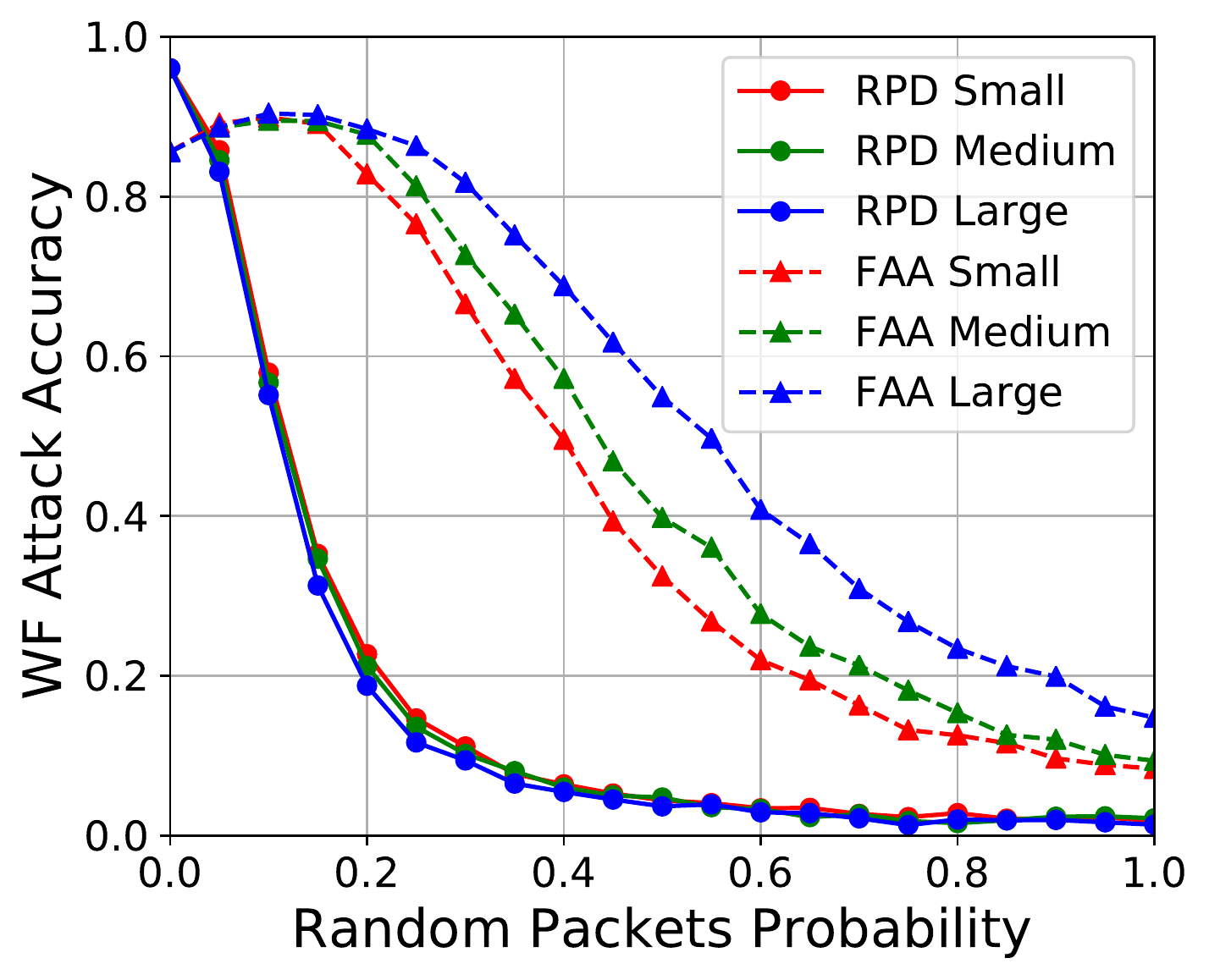}
		\caption{Microsoft Edge}
	\end{subfigure}
	\begin{subfigure}{0.24\linewidth}
		\centering
		\includegraphics[width=0.9\linewidth]{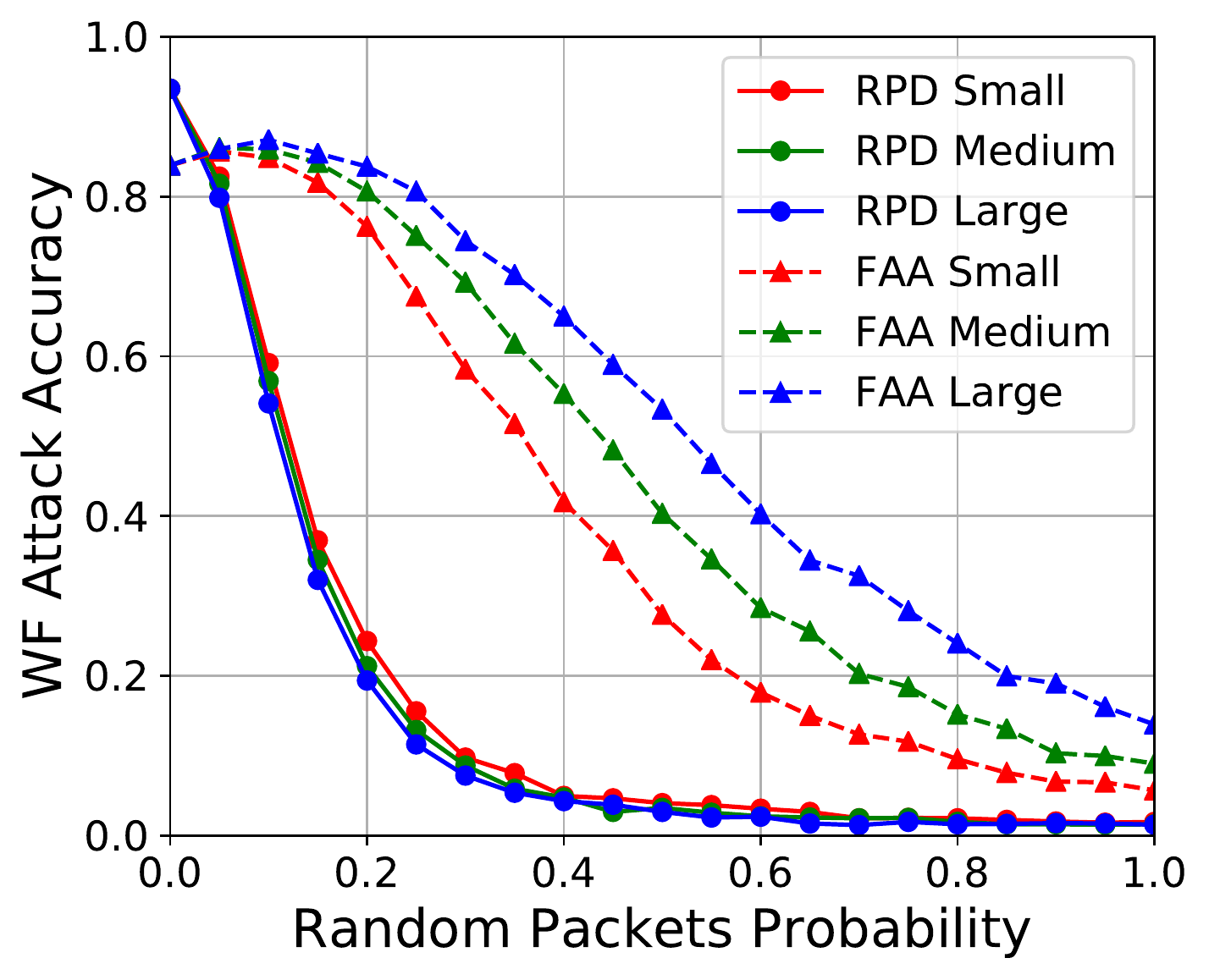}
		\caption{Google Chrome}
	\end{subfigure}
	\begin{subfigure}{0.24\linewidth}
		\centering
		\includegraphics[width=0.9\linewidth]{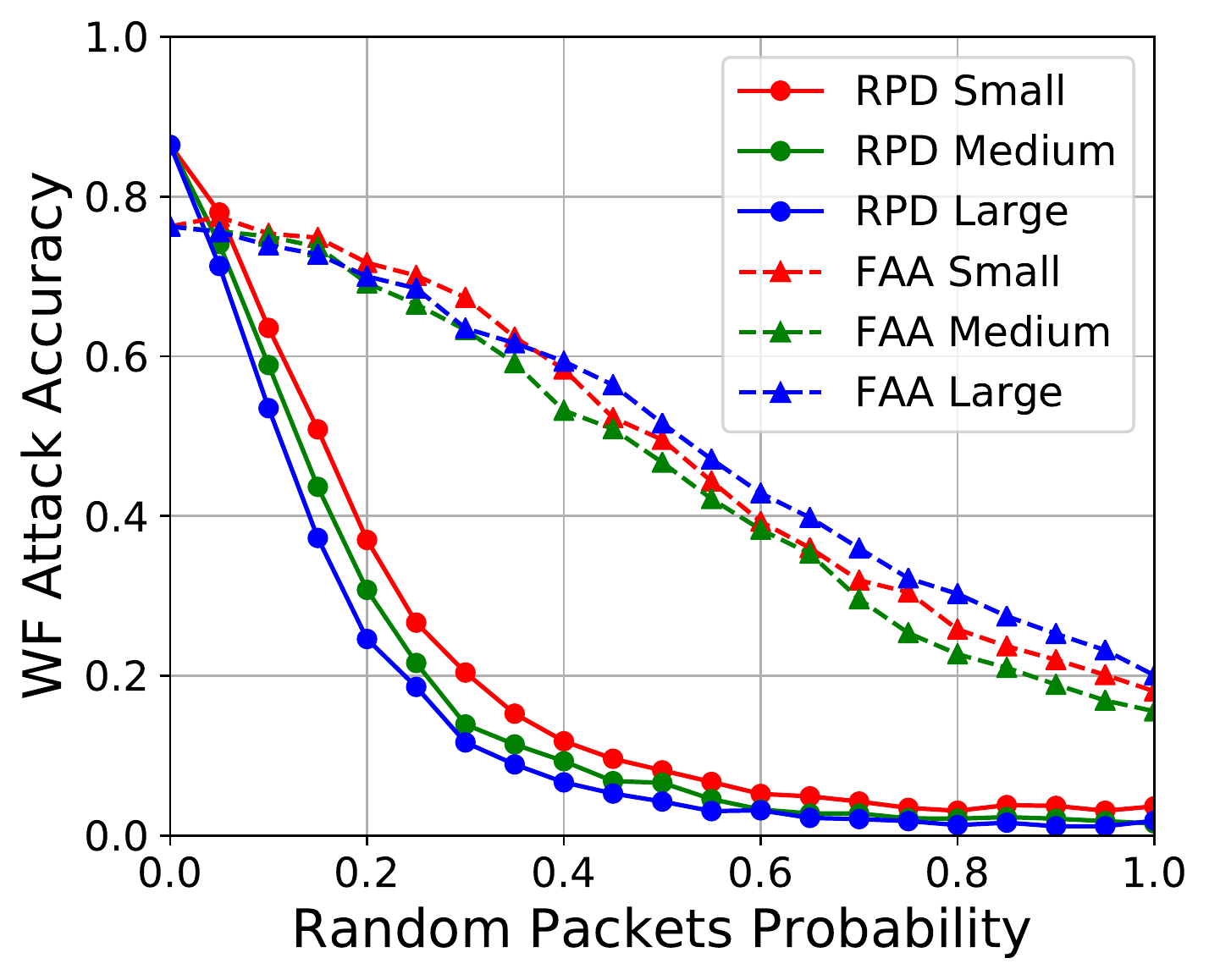}
		\caption{Mozilla Firefox}
	\end{subfigure}
  \caption{Filter Assisted Attack (FAA) on Random Packets Defense (RPD) Result}
  \label{random_filter}
\end{figure*}

\renewcommand\arraystretch{0.7}
\begin{table}[t]

  \small
  \caption{Defensive Efficiency of Random Packets Defense (RPD) and Filter Assisted Attack (FAA)}
  \label{random_de}
  \begin{center}
  \resizebox{0.95\linewidth}{!}{
  \begin{tabular}{@{}c|c|c|c|c@{}}
  \toprule
  \textbf{\ \ Browser}        & \textbf{Method} & \textbf{Small} & \textbf{Medium} & \textbf{Large\ \ } \\ \midrule
  \multirow{2}{*}{  Microsoft Edge}   & RPD Under DF    & 5.56           & 5.88            & 5.88           \\ \cmidrule(l){2-5}
                          & RPD Under FAA   & 2.33(58\%↓)    & 2.13(-64\%↓)     & 1.82(69\%↓)\ \     \\ \midrule
  \multirow{2}{*}{  Google Chrome} & RPD Under DF    & 5.88           & 6.25            & 6.25           \\ \cmidrule(l){2-5}
                          & RPD Under FAA   & 2.56(56\%↓)    & 2.17(65\%↓)     & 1.89(70\%↓)\ \     \\ \midrule
  \multirow{2}{*}{  Mozilla Firefox}& RPD Under DF    & 4.76           & 5.26            & 5.88           \\ \cmidrule(l){2-5}
                          & RPD Under FAA   & 2.04(57\%↓)    & 2.17(59\%↓)     & 2.00(66\%↓)\ \     \\ \bottomrule
  \end{tabular}
  }
  \end{center}
\end{table}

\renewcommand\arraystretch{0.7}
\begin{table}[t]
  \small
  \caption{Defensive Efficiency of List Assisted Defense (LAD) and Filter Assisted Attack (FAA)}
  \label{list_de}
  \begin{center}
  \resizebox{0.75\linewidth}{!}{
  \begin{tabular}{@{}c|c|c|c@{}}
  \toprule
  \textbf{\ \ Browser}        & \textbf{Method} & \textbf{Insert} & \textbf{Split$^1$\ \ } \\ \midrule
  \multirow{2}{*}{  Microsoft Edge}   & LAD Under DF    & 6.25            & 5.56            \\ \cmidrule(l){2-4}
                          & LAD Under FAA   & 5.26(16\%↓)     & 3.57(36\%↓)\ \      \\ \midrule
  \multirow{2}{*}{  Google Chrome} & LAD Under DF    & 6.25            & 5.88            \\ \cmidrule(l){2-4}
                          & LAD Under FAA   & 5.56(16\%↓)     & 3.57(36\%↓)\ \      \\ \midrule
  \multirow{2}{*}{  Mozilla Firefox}& LAD Under DF    & 6.25            & 5.00            \\ \cmidrule(l){2-4}
                          & LAD Under FAA   & 5.88(6\%↓)      & 3.57(29\%↓)\ \      \\ \bottomrule

  \multicolumn{4}{l}{\footnotesize $^1$ This parameter is rescaled according to packets probability.}
  \end{tabular}
  }
  \end{center}
\end{table}

\subsubsection{ROC Curve}

In order to evaluate the model more easily and comprehensively, we use closed-world and open-world assumptions. The closed-world assumption is commonly used to evaluate models, while the open-world assumption is more realistic.
In the open-world assumption, we will have four results, named
TP (True Positive, identified as positive, actually positive, correct),
TN (True Negative, identified as negative, actually negative, correct),
FP (False Positive, identified as positive, actually negative, incorrect),
FN (False Negative, identified as negative, actually positive, incorrect) to determine whether an instance is in the monitored set.
After recording the discrimination results of the whole test dataset, $\rm TPR = \frac{TP}{TP+FN}{\rm,}$ and $\rm FPR = \frac{FP}{FP+TN}{\rm}$ that describe the discrimination ability of the model are obtained.

In the process of judging whether the instance is positive, we usually set a series of thresholds between 0 and 1 and record each TPR and FPR. Finally, a ROC curve can be obtained by connecting the recorded points into lines with TPR as the ordinate and FPR as the abscissa. The more inclined the ROC curve is to the upper left corner, the higher TPR can be obtained under a certain FPR.

\subsection{Filter Assisted Attack on Random Packets Defense}

\begin{figure*}[t]
	\centering
	\begin{subfigure}{0.24\linewidth}
		\centering
		\includegraphics[width=0.9\linewidth]{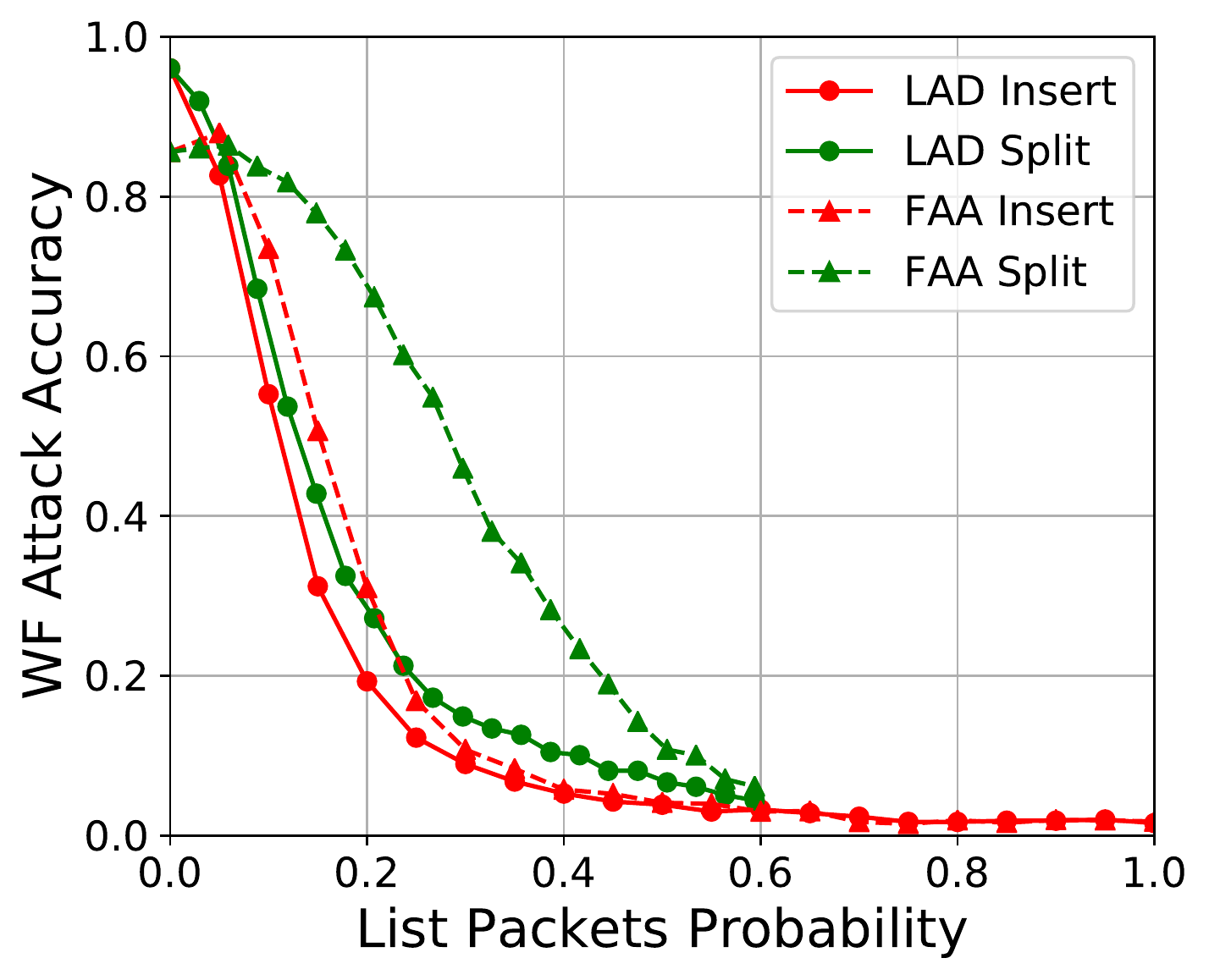}
		\caption{Microsoft Edge}
	\end{subfigure}
	\begin{subfigure}{0.24\linewidth}
		\centering
		\includegraphics[width=0.9\linewidth]{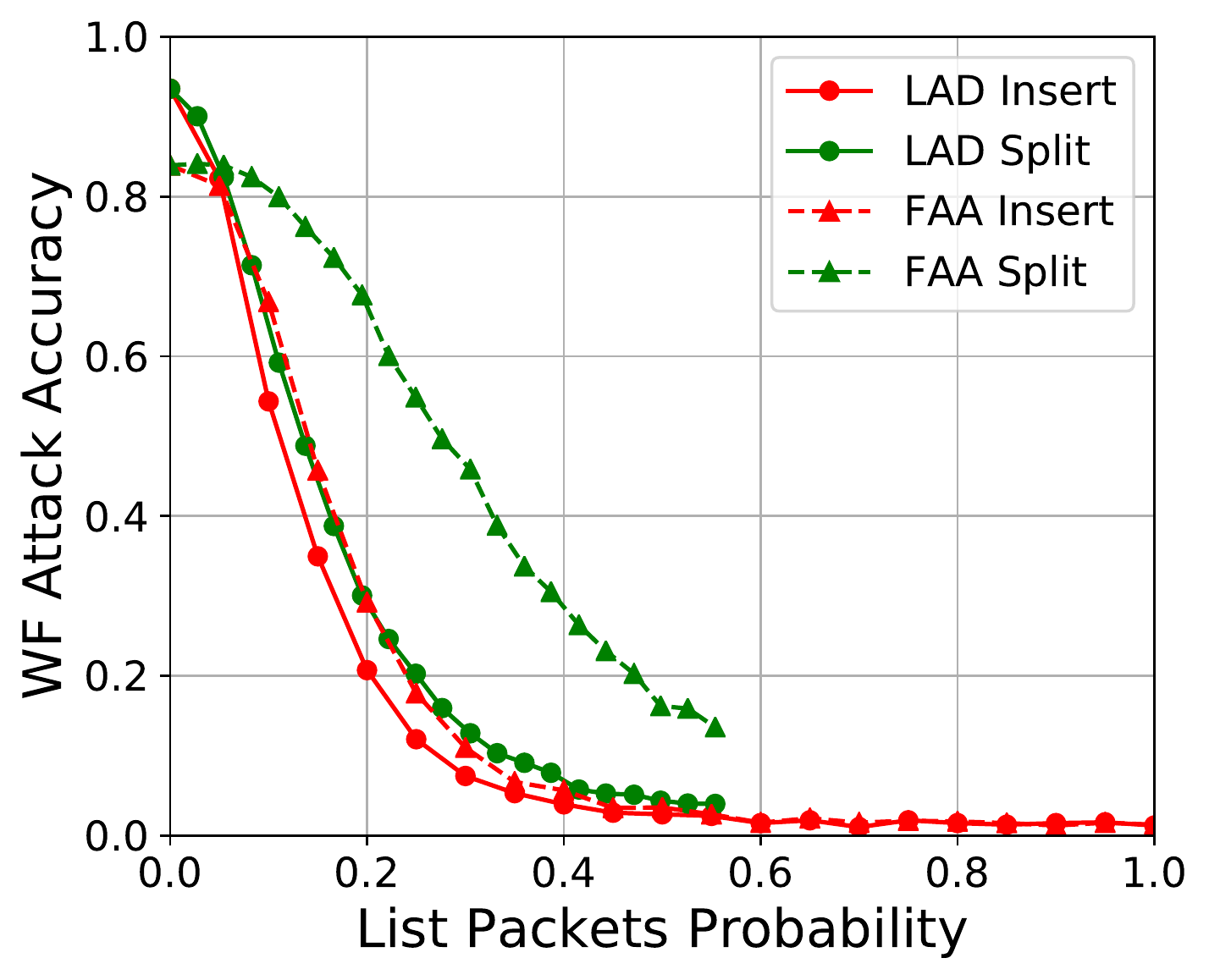}
		\caption{Google Chrome}
	\end{subfigure}
	\begin{subfigure}{0.24\linewidth}
		\centering
		\includegraphics[width=0.9\linewidth]{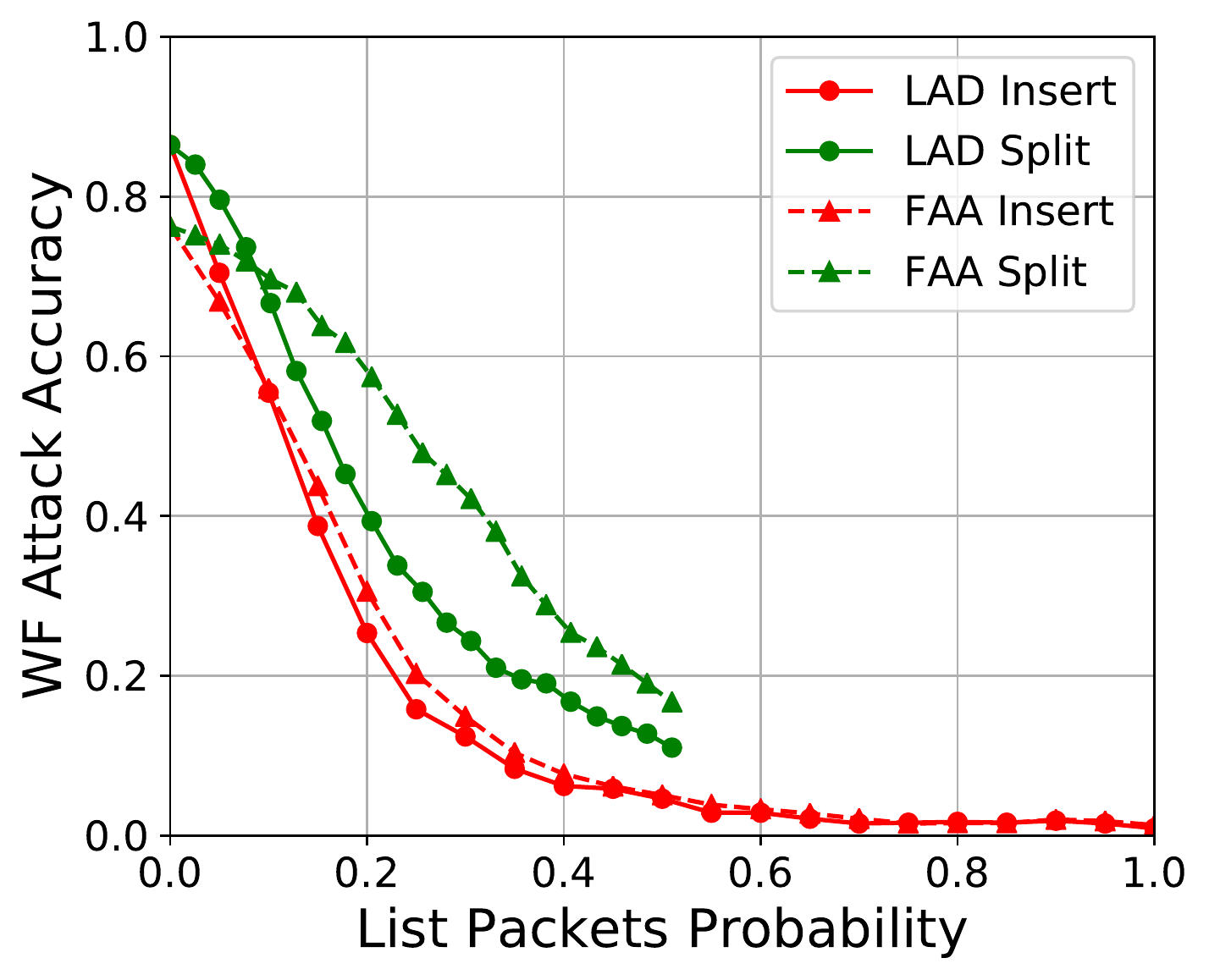}
		\caption{Mozilla Firefox}
	\end{subfigure}
  \caption{List Assisted Defense (LAD) on Filter Assisted Attack (FAA) Result}
  \label{list_filter}
\end{figure*}

\begin{figure*}[t]
	\centering
	\begin{subfigure}{0.24\linewidth}
		\centering
		\includegraphics[width=0.9\linewidth]{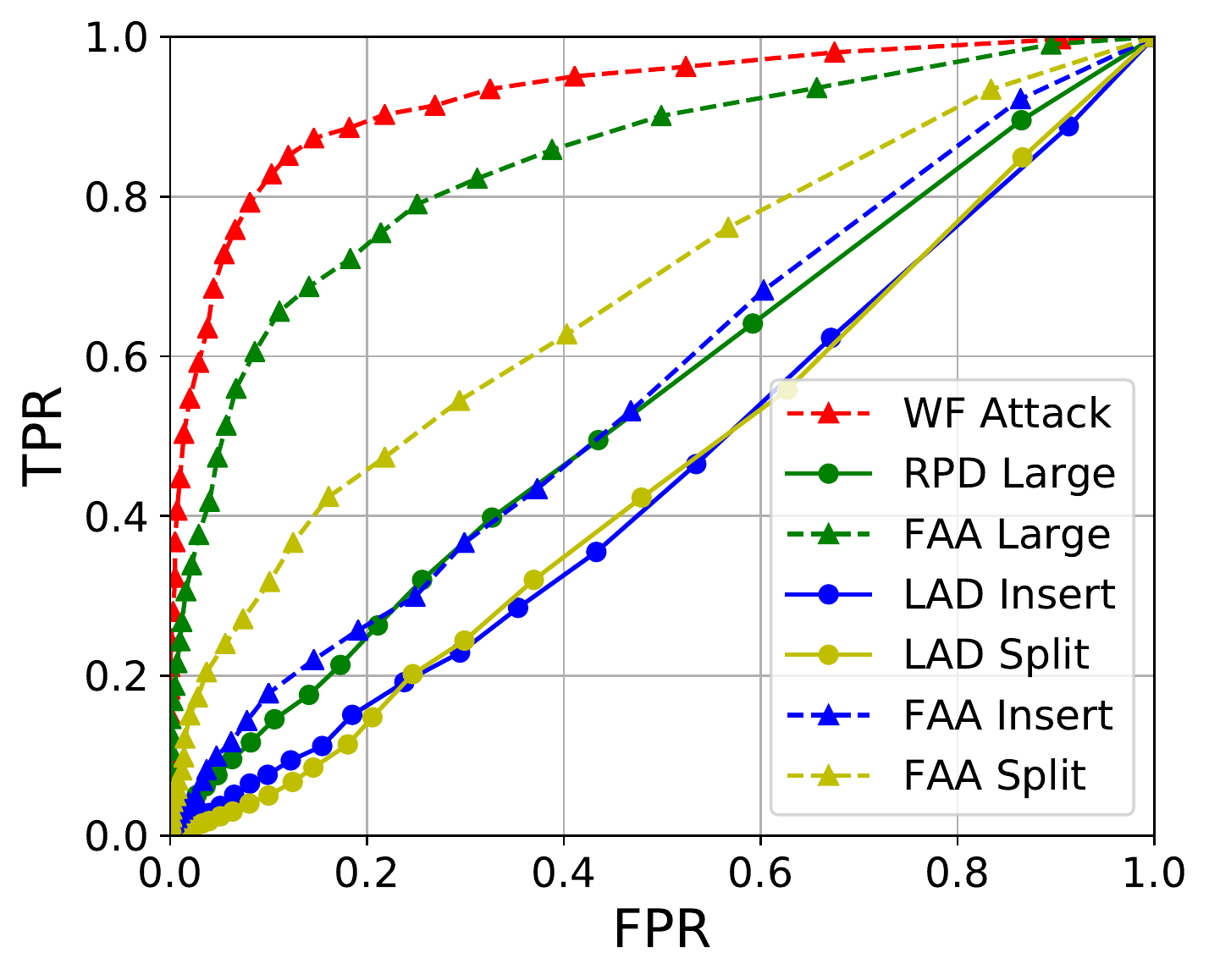}
		\caption{Microsoft Edge}
	\end{subfigure}
	\begin{subfigure}{0.24\linewidth}
		\centering
		\includegraphics[width=0.9\linewidth]{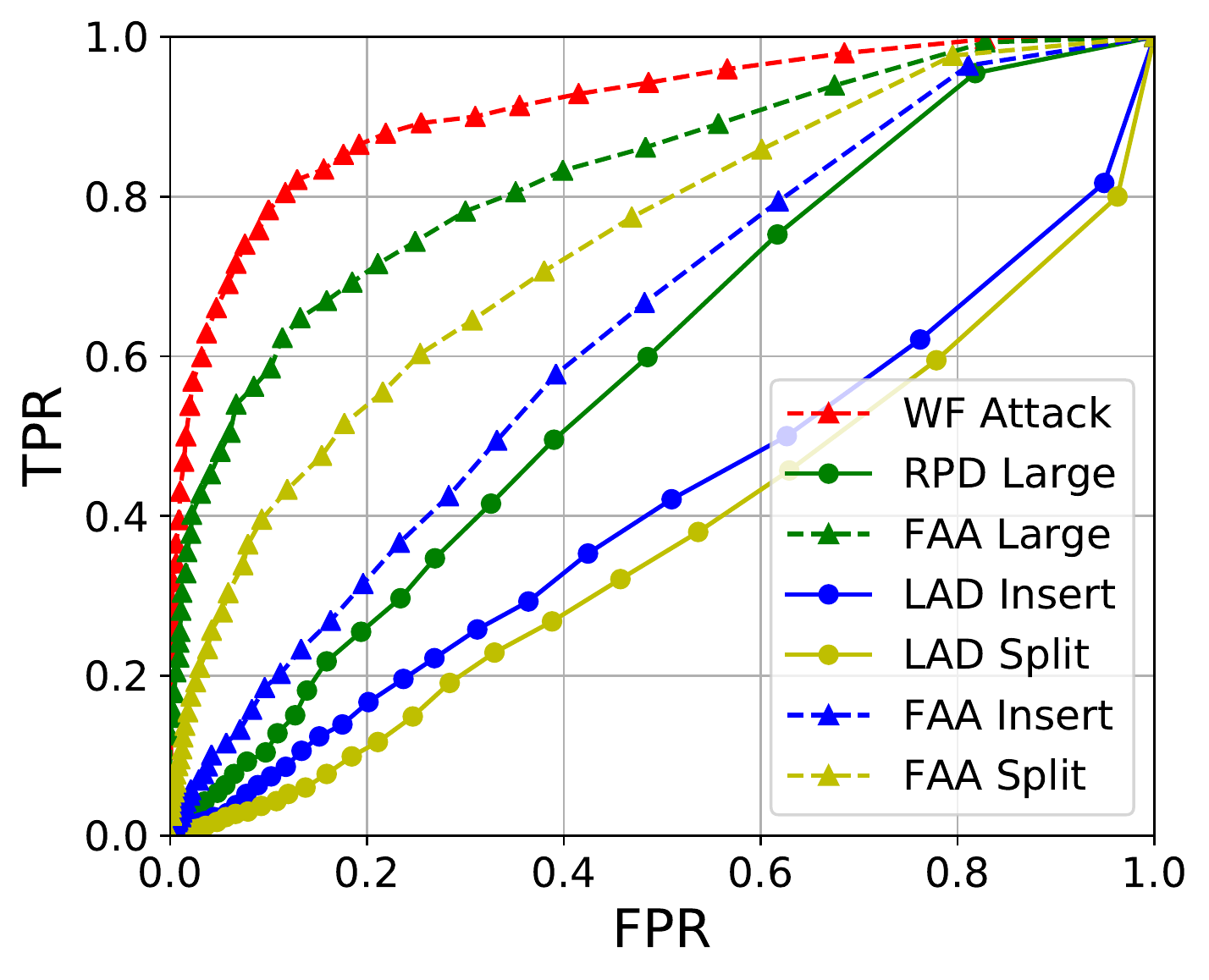}
		\caption{Google Chrome}
	\end{subfigure}
	\begin{subfigure}{0.24\linewidth}
		\centering
		\includegraphics[width=0.9\linewidth]{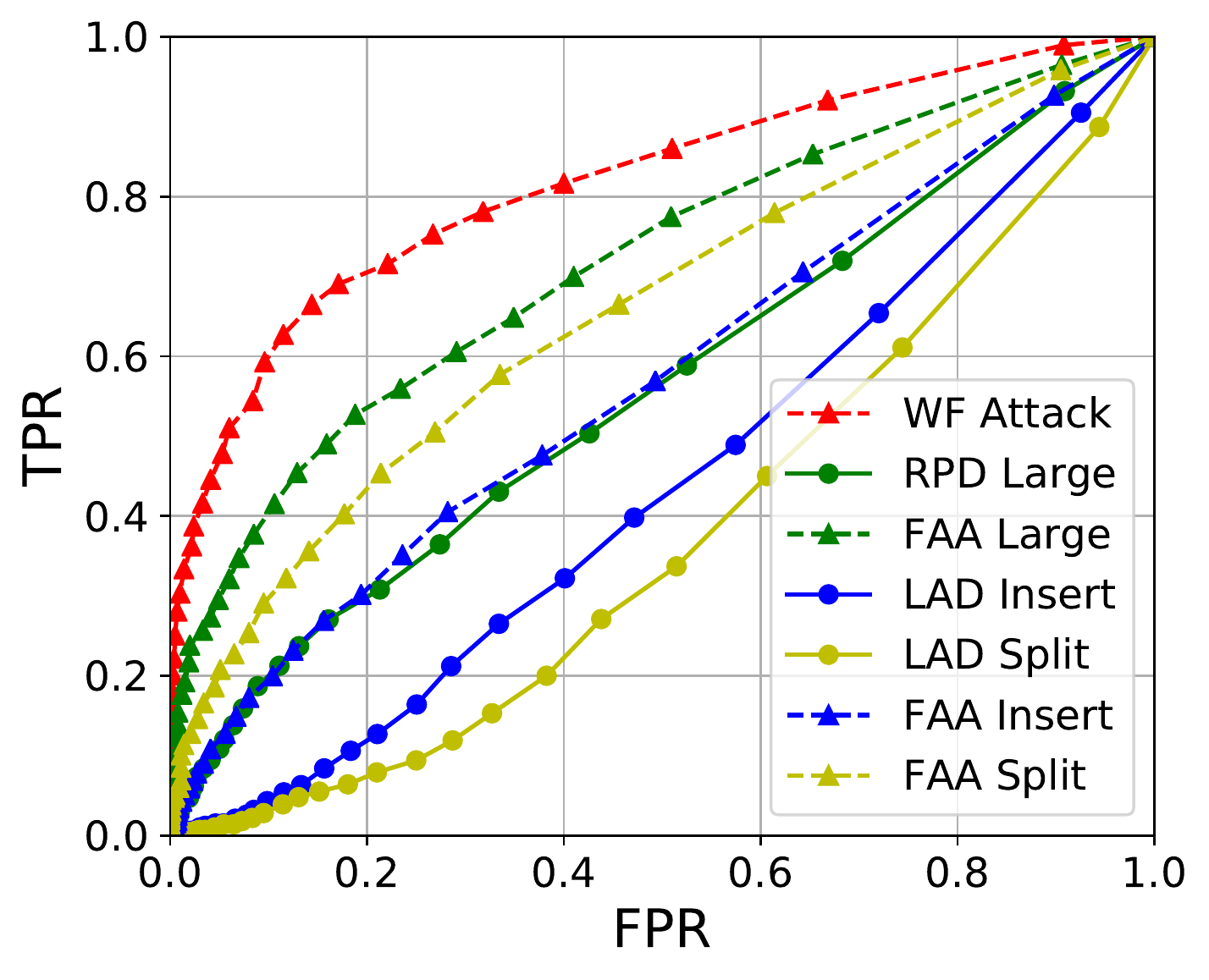}
		\caption{Mozilla Firefox}
	\end{subfigure}
  \caption{Random Packets Defense (RPD), List Assisted Defense (LAD) and Filter Assisted Attack (FAA) ROC Curve in Open World}
  \label{roc}
\end{figure*}

\begin{figure}[t]
  \centerline{\includegraphics[width=0.35\textwidth,height=4cm]{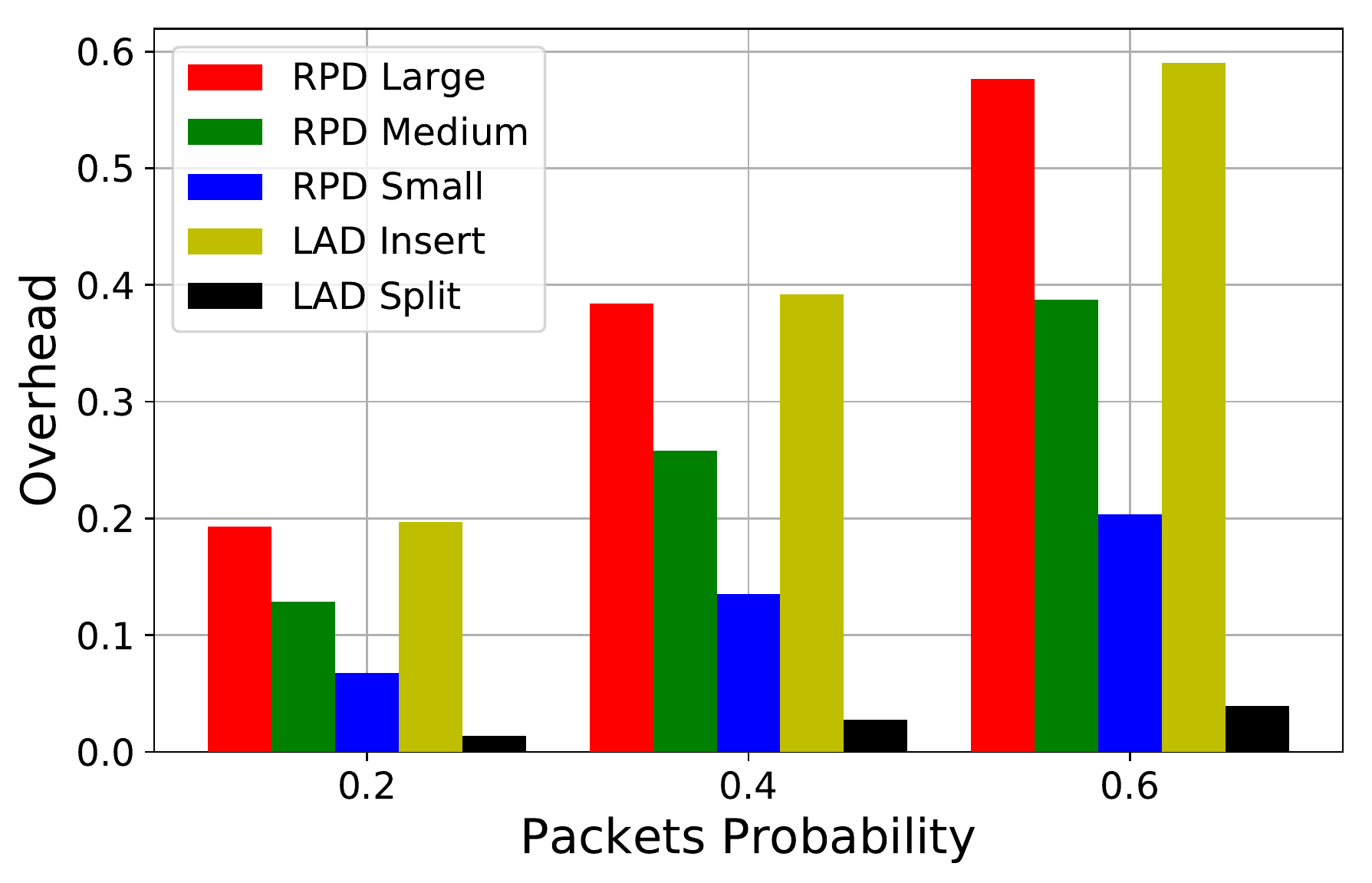}}
  \caption{Defense Traffic Overhead}
  \label{overhead}
\end{figure}

Based on the CNN model, we first conduct traffic interferences on the test dataset in terms of RPD. In addition, we select three ranges of packet size, named large packet (-1514 to 1514), medium packet (-1000 to 1000), and small packet (-500 to 500), to evaluate the influence of the size of the interference packet on the model.

The final accuracy curves are shown in the solid line of Figure~\ref{random_filter}. It can be observed that with the increase of interference packets insertion probability, the classification accuracy of DF~\cite{sirinam2018deep} has a significant decline. When the probability of random packets is greater than 20\%, the WF attack accuracy drops around 20\%. By comparing the curves of large packets, medium packets, and small packets, we can also find that the size range of packets has a limited impact on the accuracy degradation. Therefore, in RPD, we can use as many small packets as possible to fill in, to achieve the lowest network overhead with negligible performance losses.

Then, we use the FAA method to carry out enhanced attacks on the above RPD traffic, and the accuracy curves are shown in the dashed line in Figure~\ref{random_filter}. It can be seen that after FAA is applied, the decrease in recognition accuracy significantly slows down. In addition, the DE of RPD and DE under FAA are provided in Table~\ref{random_de}. We can see that the FAA significantly reduces the defense efficiency of RPD, where the drops are from 70\% to 56\% for large and small injected packet sizes, respectively.

\subsection{List Assisted Defense on Filter Assisted Attack}

In this experiment, we interfere with the traffic according to the method of LAD, and the results are provided in Figure~\ref{list_filter} and Table~\ref{list_de}. The solid line represents the effect of LAD while the dashed line is FAA. It can be seen that with the increase of packet insertion probability, the classification accuracy drops significantly. Among them, the LAD based on packet insertion has the best defense effect. Even if the proposed FAA is applied, the accuracy decline is not basically slowed down (e.g., the drop of DE is smaller than 16\%). However, the LAD based on packet splitting has a slightly slower precision decline under the FAA (e.g., the drop of DE is about 30\%). The reason is that packet splitting can not guarantee the size of the two newly generated packets both in the list, so one has a relatively high probability of being detected. Therefore, its resistance effect against FAA is slightly worse than that of LAD based on packet insertion.

\subsection{Defense Traffic Overhead}

In Figure~\ref{overhead}, we provide the traffic overhead of RPD and LAD. We select 20\%, 40\%, and 60\% packet probabilities, respectively,  as test points, and compare the overhead of defense measures. As can  be observed, the overhead of large, medium, and small packets of RPD decreases proportionally. Since the packet insertion needs to keep the probability distribution of the original traffic, the range of the inserted packets is the same as the large packets of RPD, so its overhead is the same as that of the large packets of RPD. However, the interference of LAD based on packets splitting comes from the traffic itself, it only needs to insert extra packet headers, so its overhead is far lower than other defense methods. In general, LAD based on packet insertion has the best defensive performance with a relatively high overhead, while LAD based on packet splitting is better and balanced in both defense ability and overhead.

\subsection{Open World Evaluation}

In this subsection, we evaluate the ROC curves of RPD and LAD under FAA. The results are shown in Figure~\ref{roc}. By comparison, we can find that the ROC of the WF attack is the highest without any defense measures. When we apply RPD and LAD, ROC decreases significantly. However, when we apply FAA to enhance the attack, RPD will be defeated, while LAD based on packet insertion still has a satisfactory performance, which is consistent with the results in the closed-world dataset.

\section{Conclusion}

In this paper, we have first studied the influence of random noise on WF attacks based on traditional random packet defense. Then a filter-assisted attack has been proposed to mitigate the noise and enhance the efficiency of the attack. Finally, we have investigated a  list-assisted defense in dealing with the filter-assisted attack. From our experimental results, we observe that the list-assisted defense based on packets insertion achieves the best performance and can resist the filter-assisted attack, but its network overhead is close to that of random large packets, while the list-assisted defense based on packet splitting strikes a balance in both performance and overhead, which is proved to be an efficient and feasible method.

%
\begin{acks}
This work is partially supported by the National Key R\&D Program of China (2021YFB3100700), the National Natural Science Foundation of China (62272228, 62002170, 62076125, U20B2049, U20B2050, U21A20467), Shenzhen Science and Technology Program (Grant No.JCYJ20210324134408023), Youth Foundation Project of Zhejiang Lab (No. K2023PD0AA01), and Research Initiation Project of Zhejiang Lab (No. 2022PD0AC02).
\end{acks}

\clearpage

\nocite{*}
\bibliographystyle{ACM-Reference-Format}
\bibliography{sample-base}

\end{document}